\documentclass[reqno,12pt]{article}
\usepackage{amsmath,amsfonts,amssymb,amsthm,amstext,amscd,eucal,xcolor}
\usepackage[all]{xy}
\usepackage{amsmath, amssymb}
\usepackage{mathtools}
\newcommand{\mathsym}[1]{{}} 
\usepackage[colorlinks=true,
            linkcolor=blue,
            urlcolor=blue,
            citecolor=blue]{hyperref}
\usepackage{enumerate}
\usepackage{slashed}

\usepackage{listings}
\DeclareMathAlphabet{\pazocal}{OMS}{zplm}{m}{n}
\usepackage{graphicx}
\usepackage{amsmath} \usepackage{amssymb}
\usepackage{bm}
\usepackage{cite,hyperref}
\usepackage{mathtools}
\usepackage{amsmath} \usepackage{amssymb}
\usepackage{bm}
\usepackage{graphicx}
\usepackage[active]{srcltx}
 \usepackage{color}
\definecolor{DeepPink}{rgb}{1.00,0.08,0.58}
\definecolor{DeepPink2}{rgb}{0.93,0.07,0.54}
\makeatletter \@addtoreset{equation}{section}

\makeatletter\renewcommand\section{\@startsection {section}{1}{\z@}%
                                   {-3.5ex \@plus -1ex \@minus -.2ex}
                                   {2.3ex \@plus.2ex}%
                                   {\normalfont\large\bfseries}}
\renewcommand\subsection{\@startsection{subsection}{2}{\z@}%
                                     {-3.25ex\@plus -1ex \@minus -.2ex}%
                                     {1.5ex \@plus .2ex}%
                                     {\normalfont\bfseries}}

\DeclareMathAlphabet{\mathcal}{OMS}{cmsy}{b}{n}

\makeatother

\newcommand{\email}[1]{\footnote{E-mail: \href{mailto:#1}{#1}}}

\begin{document}

\title{\bf\Large{ Direct single-shot phase difference retrieval of two wavelength in off-axis digital holography}}

\author{\textbf{M. Amani And M.~Dashtdar\email{m-dashtdar@sbu.ac.ir}}\\\\
\textit{\small   Department of Physics, Shahid Beheshti University, 1983969411, Tehran, Iran}
}
\maketitle
\date{}

\begin{abstract}
A method for direct phase difference reconstruction using single-shot dual-wavelength off-axis digital holography is presented. This approach enables direct imaging of samples with high steps without the need to reconstruct phase images at each individual wavelength. As the dual wavelengths in the reference and object arms pass through a common path in this configuration, single-wavelength arrangements can be applied. Due to the unique capability of the presented method, a sodium-vapor lamp source has been utilized to obtain two closely spaced wavelengths ($\lambda_1 = 589 \ nm$ and $\lambda_2 = 589.6 \ nm$), with in a synthetic wavelength of $\Lambda = 578.8\ \mu m$ in the Michelson configuration.
To evaluate the validity of the method, the height of an air wedge measured using the proposed approach has been compared with the result obtained from phase unwrapping in the single-wavelength method. The capability of the proposed technique to image samples with high step structures is further demonstrated by measuring a $30\ \mu m$ step height and a glass plate with a thickness of approximately $140\ \mu m$.
\end{abstract}


\setcounter{footnote}{0}
\renewcommand{\baselinestretch}{1.05}  



\section{Introduction}
Digital holography (DH) is an imaging technique that uses digital image sensors such as a charge-coupled device (CCD) or a complementary metal-oxide semiconductor (CMOS) for hologram recording \cite{schnars2015digital,rastogi2013holographic,goodman1967digital,cuche1999digital,pomarico1995digital,kreis2006handbook,kamilov2015learning,ren2018learning}. For this reason, in DH, unlike traditional holography, both the intensity and phase details of a recorded holographic wavefront are directly acquired during the numerical reconstruction process, eliminating the need for chemical processing. Because of these advantages, DH has many applications in the fields of three-dimensional (3D) imaging, microscopy, metrology, display technology, and data storage \cite{frauel2006three,marquet2005digital,kim2010principles,bhaduri2014diffraction,kreis2006handbook,geng2013three,coufal2000holographic,osten2014recent,park2018quantitative,popescu2011quantitative,park2010measurement,park2010metabolic,wu2018extended}. 

In single-wavelength DH, the reconstructed phase map is often limited by $2\pi$ phase ambiguities due to the constrained illumination wavelength \cite{cuche1999digital}.  Phase unwrapping algorithms are used to remove this phase ambiguity and obtain a continuous phase map of the object \cite{takeda1982fourier}. These algorithms fail for specimens with step height greater than one wavelength and noisy experimental conditions. In addition, such unwrapping methods that are often time-consuming are inadequate for real time monitoring of fast processes.
To solve these drawbacks and extend the range of phase measurement, dual-wavelength DH with a large synthetic wavelength is introduced. Since the synthetic wavelength obtained from two different wavelengths is longer than each of the individual wavelengths, the measurement range in DH is extended. One of the dual-wavelength DH methods involves recording multiple interferograms sequentially  \cite{wada2008multiple,wada2008large,polhemus1973two}. Although this method can greatly extend the phase measurement range, it is not single shot and therefore not suitable for dynamic processes.
To achieve real-time measurement, interferograms of two wavelengths are simultaneously recorded. Two interferograms can be recorded simultaneously through different color channels of a color Bayer-mosaic camera \cite{min2012dual,min2018optical,rinehart2010simultaneous}. As the peak spectral response characteristics of Bayer filters in different channels limit the use of two wavelengths, the large synthetic wavelengths cannot be created. Additionally, spectral overlap between different channels causes errors in phase measurements. Dual-wavelength DH utilizing two monochrome cameras eliminates the limitations associated with using two wavelengths in a color Bayer-mosaic camera. \cite{ishigaki2021height,ibrahim2019simultaneous,li2023dual}. However, this approach demands precise individual alignment of both cameras, making the system setup technically complicated. Eventually, the large setup of this method results in reduced temporal stability, as environmental disturbances like mechanical fluctuations can affect each beam differently.
An alternative method for simultaneous recording of two holograms involves off-axis dual-wavelength DH with a monochrome camera. In this case, the spatial frequency spectrum of the object carried by two wavelengths propagated in the same directions is separated. This separation enables the reconstruction of the two phase maps separately. However, it is difficult to separate the frequency spectrum of the object for closely related wavelengths due to the overlap of frequencies \cite{di2016dual,kumar2020single,nourzadeh2023phase}. To solve this problem, orthogonal interference fringes are obtained by adjusting the propagation direction of different wavelengths. In this case, the two object spectra in frequency space are separated in different directions due to crossed fringes in the interferogram \cite{kuhn2007real,wu2024convenient,guo2017compact,liu2017simultaneous,huang2021real}. Although this method allows for a larger synthetic wavelength and has a less complicated arrangement compared to using two monochrome cameras, it has two disadvantages. First, due to the separate reconstruction of phases, phase noises corresponding to each of the object waves are amplified in the synthetic wavelength. Second, because two reference beams with independently adjustable propagation directions must be separately aligned, this method cannot be used in common-path configurations characterized by high stability and low mechanical fluctuations \cite{ebrahimi2018stable,s2019common,calabuig2014common,ebrahimi2021lens,ebrahimi2019quantitative,bhaduri2014diffraction}.

In this letter, firstly, a method based on the direct phase difference measurement using dual-wavelength off-axis digital holography is introduced. In this method, the phases of the two wavelengths are not reconstructed separately. Thus, spatial frequency spectrum separation of the object in Fourier space is not required, and the limitation in choosing two closely related wavelengths is eliminated, allowing for the creation of larger synthetic wavelengths. Moreover, due to the common path of both object and reference beams for each two wavelength, this method can be used in whole conventional single-wavelength setups.

Secondly, in order to verify the proposed method, a sodium-vapor lamp with two spectral lines, $\lambda_1=589\ nm$ and $\lambda_2=589.6\ nm$, with a synthetic wavelength $\Lambda=578.8 \ \mu m$ is used as a source.  Our method uniquely enables the use of a single source with two spectral lines, a capability not found in any other system. Utilizing a sodium-vapor lamp offers multiple advantages compared to using two lasers. There is‌ a coupling between two wavelengths due to emission from the same source. In this case, the effect of fluctuations due to the use of two different laser sources will be eliminated and environmental noise will be reduced. Also, sodium lamp has less speckle noise due to lower coherence than laser \cite{jeon2016dual}.

\section{Theory}
In dual-wavelength digital holography, the intensity distribution of the recorded hologram is given by
\vspace{-3pt}
\begin{equation}
	\begin{aligned}
		I(x,y) =& |O_1|^2+|R_1|^2+|O_2|^2+|R_2|^2 \\+ & O_1 R_1^*+ O_2 R_2^*+O_1^* R_1+O_2^* R_2 ,
		\label{eq1}
	\end{aligned}
\end{equation}
Here, $O_1$ and $R_1$ are object and reference beams for $\lambda_1$, respectively; $O_2$ and $R_2$ are object and reference beams for $\lambda_2$, respectively;  The first four terms in Eq. (\ref{eq1}) represent the zero-diffraction order, and the next four terms represent the interference terms between the object and reference beams. On the Fourier plane of Eq. (\ref{eq1}), due to the overlap of spatial frequencies in the Fourier domain, the $\pm1$ orders of frequency spectra for $\lambda_1$ and $\lambda_2$ are not separated. In this case, one order of the frequency spectrum of the object beam for $\lambda_1$ and $\lambda_2$ can be isolated with a filtering mask. After the Inverse Fourier transform, the object beam is given by
\vspace{-4pt}
\begin{equation}
	\begin{aligned}
		U(x,y) = & \ O_1 R_1^*+ O_2 R_2^*\\=&|O_1(x,y)||R_1(x,y)|e^{-i\varphi_1}+|O_2(x,y)||R_2(x,y)|e^{-i\varphi_2},
		\label{eq33}
	\end{aligned}
\end{equation} 

Where, $\varphi_1$ and $\varphi_2$ denote phase difference between object and reference beams for $\lambda_1$ and $\lambda_2$, respectively. For $|O_1|=|O_2|=|O|, |R_1|=|R_2|=|R|$, we have 
\begin{equation}
	\begin{aligned}
		U(x,y) =&\  |O(x,y)| |R(x,y)| \Bigg[\Bigg( \cos\Big(\varphi_1(x,y)\Big) + \cos\Big(\varphi_2(x,y) \Big)\Bigg)\\-&i\Bigg(\sin\Big(\varphi_1(x,y) \Big) + \sin\Big(\varphi_2(x,y)\Big)\Bigg)\Bigg].
		\label{eq44}
	\end{aligned}
\end{equation}
After some deduction, we can write 
\vspace{-3pt}
\begin{equation}
	\begin{aligned}
		U(x,y) = & \ 2 |O(x,y)| |R(x,y)|\textbf{.} \Bigg[ \Bigg( \cos\left({\bar{\phi}}\right) \cos\left(\frac{\Delta\phi(x,y)}{2}\right) \Bigg) 
		\\-& i \Bigg( \sin\left({\bar{\phi}}\right) \cos\left(\frac{\Delta\phi(x,y)}{2}\right) \Bigg) \Bigg],
		\label{eq55}
	\end{aligned}
\end{equation}

Here, $\Delta\phi(x,y)=\varphi_1(x,y)-\varphi_2(x,y)$ is the phase difference between the dual-wavelength, and $\bar{\phi}=[\varphi_1(x,y) + \varphi_2(x,y)]/2$ is the mean phase of the dual-wavelength.

Considering 
$T(x,y)=\sqrt{\left| U(x,y)\right| ^2}$ and $A\left(x,y\right)=2|R\left(x,y\right)||O\left(x,y\right)|$, phase difference for the dual-wavelength approach can be obtained  

\vspace{-5pt}
\begin{equation}
	\Delta\phi(x,y)=2cos^{-1}\Bigg(\frac{T\left(x,y\right)}{A\left(x,y\right)}\Bigg).
	\label{eq6}
\end{equation}

\section{Experimental Details}
\subsection{Optical Experimental System}
To verify the proposed method experimentally, a conventional Michelson interferometer is used. Since there is no need to separate the two spectral lines in the proposed method, a single source can be used in the setup. Here, to produce two wavelengths close to each other, a sodium-vapor lamp with wavelengths $\lambda_1=589\ nm$ and $\lambda_2=589.6\ nm$ is used as a source. These two wavelengths are coupled together, which eliminates fluctuations related to using two separate lasers. The colimated sodium-vapor light after passing through a narrow bandpass filter with the
central wavelength of 589 nm and the spectral bandwidth of 10 nm to eliminate other wavelengths, is split by the beam splitter and reflected from two mirrors.  One mirror (movable mirror) is mounted on a movable stage capable of displacement with an accuracy of $2\ \mu m$. The other mirror (fixed mirror) remains stationary and is tilted to achieve an off-axis configuration. The reflected beams are recombined at the beam splitter and interfere. The resulting interference fringes after passing through a converging lens are captured by a CMOS sensor (Point Gray BFLY-U3-23S6M-C, with an 8-bit dynamic range and a pixel pitch of $5.86\ \mu m$).

\begin{figure}[h]
	\centering
	\centering{\includegraphics[width=0.6\textwidth]{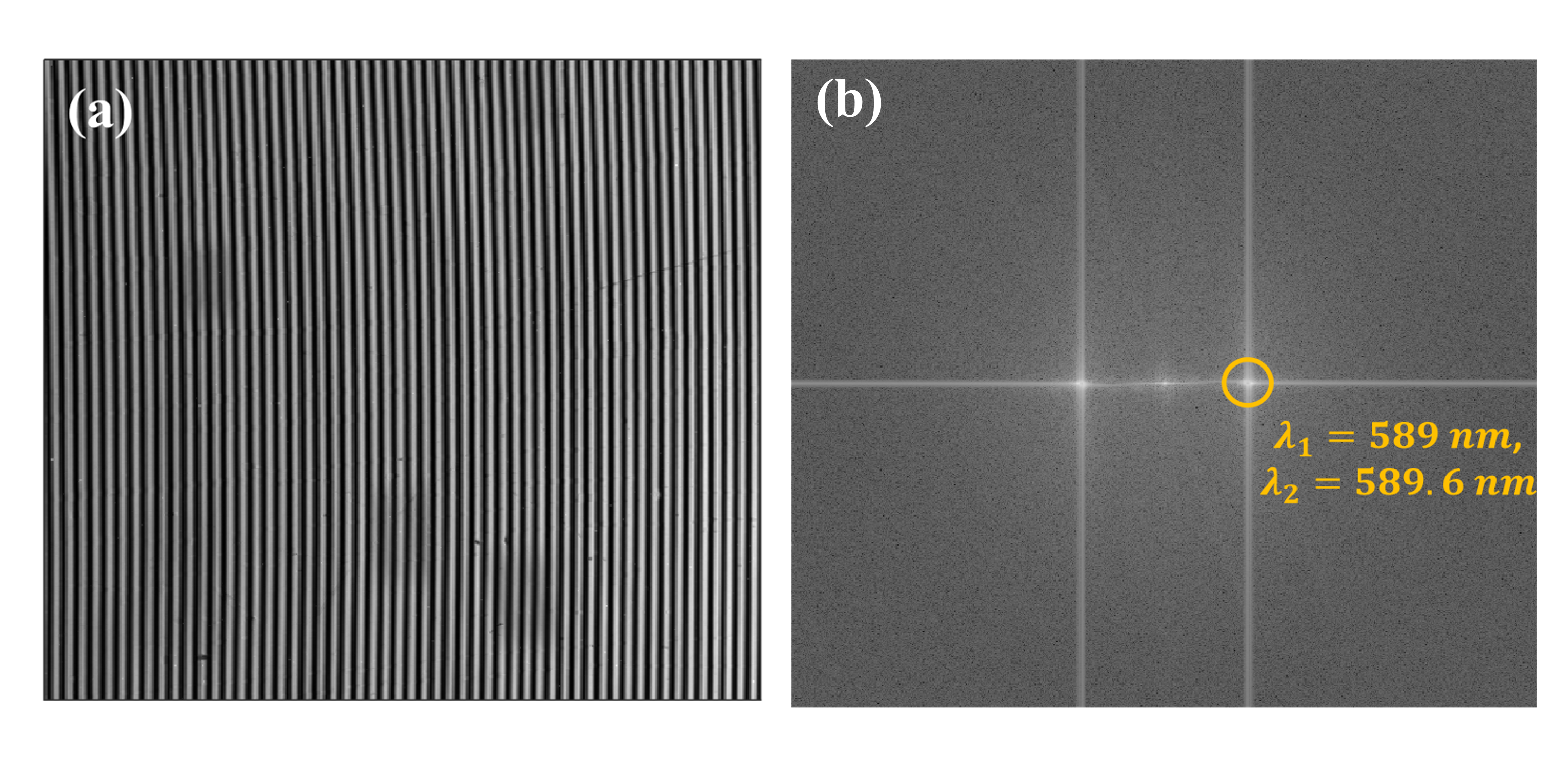}}
	\vspace*{-5mm}
	\caption{(a) Recorded hologram. (b) Fourier spectra, in which frequency spectra for $\lambda_1$ and $\lambda_2$ in the yellow circle are not separated.}
	\label{Pic1}
\end{figure}

Figures \ref{Pic1}(a) and  \ref{Pic1}(b) show the recorded dual-wavelength digital hologram and its Fourier spectrum, respectively. The $+1$ order spectra for $\lambda_1$ and $\lambda_2$ are marked with a yellow circle in the Fourier spectrum. Due to the overlap of the spatial frequencies of $\lambda_1$ and $\lambda_2$, the object spectra in frequency space are not separate.

\subsection{Calibration of A(x,y)}
\begin{figure}[h!]
	\centering
	\centering{\includegraphics[width=0.7\textwidth]{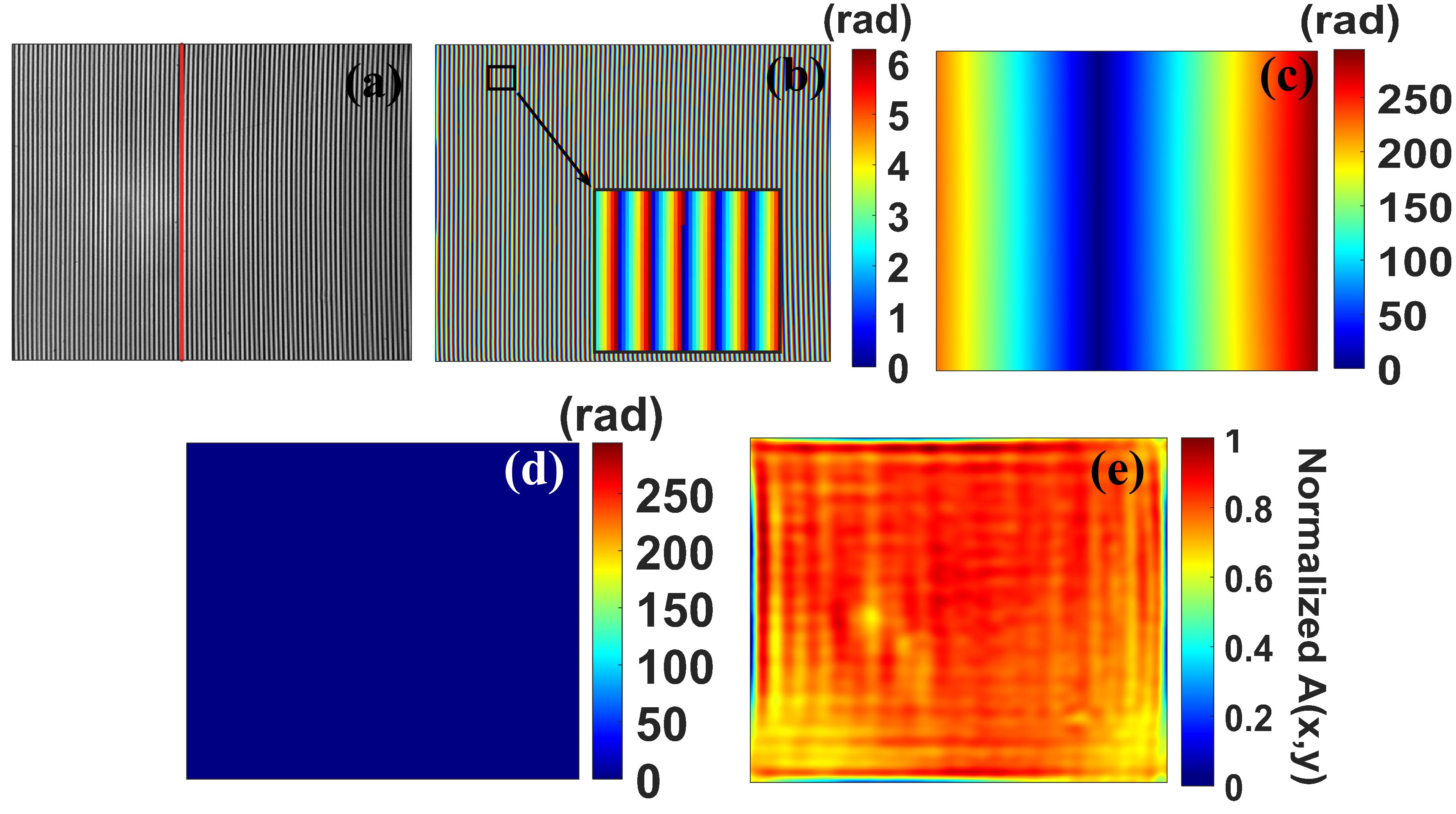}}
	\vspace*{-3mm}
	\caption{(a) Recorded background hologram; the red line is where $\Delta\phi=0$. (b) Wrapped Phase, and (c) Unwrapped phase for $\lambda=589.3 \ nm$ with single wavelength method. (d) Reconstructed phase for  $\Lambda=578.8 \ \mu m$. (e) Normalized $A\left(x,y\right)$. }
	\label{Pic2}
\end{figure}

To retrieve the phase difference from Eq. \ref{eq6}, \( A(x, y) \) should be determined. It can be obtained by directly measuring the intensity distribution of the object and reference beams. However, for single-shot measurements, it can be determinad by capturing a background hologram with the condition that the equal optical path is in the field of view. If the intensity distribution is uniform, \( A(x, y) \) will be constant across the entire hologram and can be determined when  \(\Delta \phi(x, y) = 0\) (i.e. it is equal to maximum of \(T(x, y)\)). In the situation that the intensity has not uniform spatial distribution, \( A(x, y) \) can be determined by considering single-wavelength phase retrieval algorithm for $\lambda=589.3\ nm$. As shown in Fig \ref{Pic2}(a), we first record a background hologram that included \(\Delta \phi(x, y) = 0\), which is indicated by the red line.
Then, the single-wavelength phase of the recorded hologram is reconstructed using the angular spectrum method \cite{anand2010real}.  By determining the location of 
$\Delta\phi(x,y)=0$, the continuous phase of a single wavelength can be easily obtained using the phase unwrapping method \cite{takeda1982fourier}. Figure \ref{Pic2}(b) and \ref{Pic2}(c) show  the wrapped and unwrapped phase map, respectively. The differential phase of the synthetic wavelength can be written as  	$\Delta\phi_\Lambda(x,y)=\Delta\phi_\lambda(x,y)\lambda/\Lambda$, where $\Delta\phi_\lambda$ is the retrieved unwrapped phase for $\lambda=589.3\ nm$; $\Lambda=\lambda_1\lambda_2/\left|\lambda_1-\lambda_2\right|$  denotes synthetic wavelength ($\Lambda=578.8\ \mu m$). $\Delta\phi_\Lambda(x,y)$ shows in Fig \ref{Pic2}(d). Eventually, we can obtain  $A\left(x,y\right)$ for each pixel by putting  $\Delta\phi_\Lambda(x,y)$ in Eq. (\ref{eq6}), as shown in Fig. \ref{Pic2}(e).

\vspace{-3pt}

\begin{figure}[h!]
	\centering
	\includegraphics[width=0.47\textwidth]{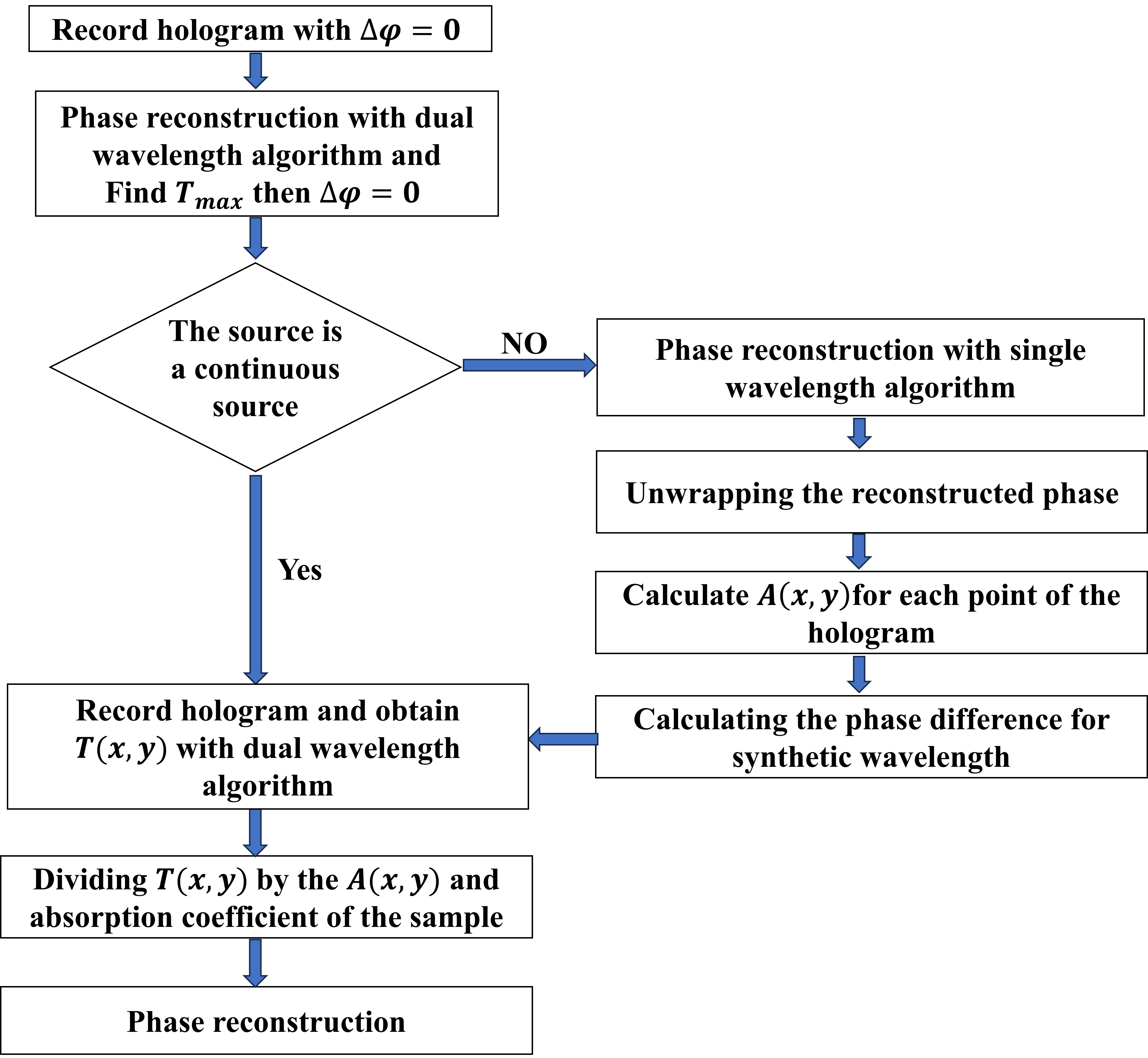}
	
	\caption{The whole process of the proposed dual-wavelength algorithm.}
	\label{Pic3}
\end{figure}

After determination of \( A(x,y) \) for background hologram, since \( A(x,y) \) of the sample hologram is equal to the product of the background and the absorption coefficient of the sample, the absorption coefficient can be determined from the zero order diffraction in the Fourier transform of the hologram. Therefore, the phase difference of two wavelength can be directly determined from Eq. \ref{eq6}.  Figure \ref{Pic3} shows the entire process of the proposed dual-wavelength algorithm.

\section{Results And Discussion}
To verify the proposed method, we created an air wedge in the Michelson interferometer by adjusting the reference mirror angle. We adjusted the wedge angle in such a way that a zero phase difference was positioned within the field of view, thereby enabling the use of the single-wavelength method.
Then we reconstructed and measured the maximum height of the air wedge within the field of view using a single-wavelength and our proposed dual-wavelength methods.

\begin{figure}[h!]
	\centering
	\includegraphics[width=0.7\textwidth]{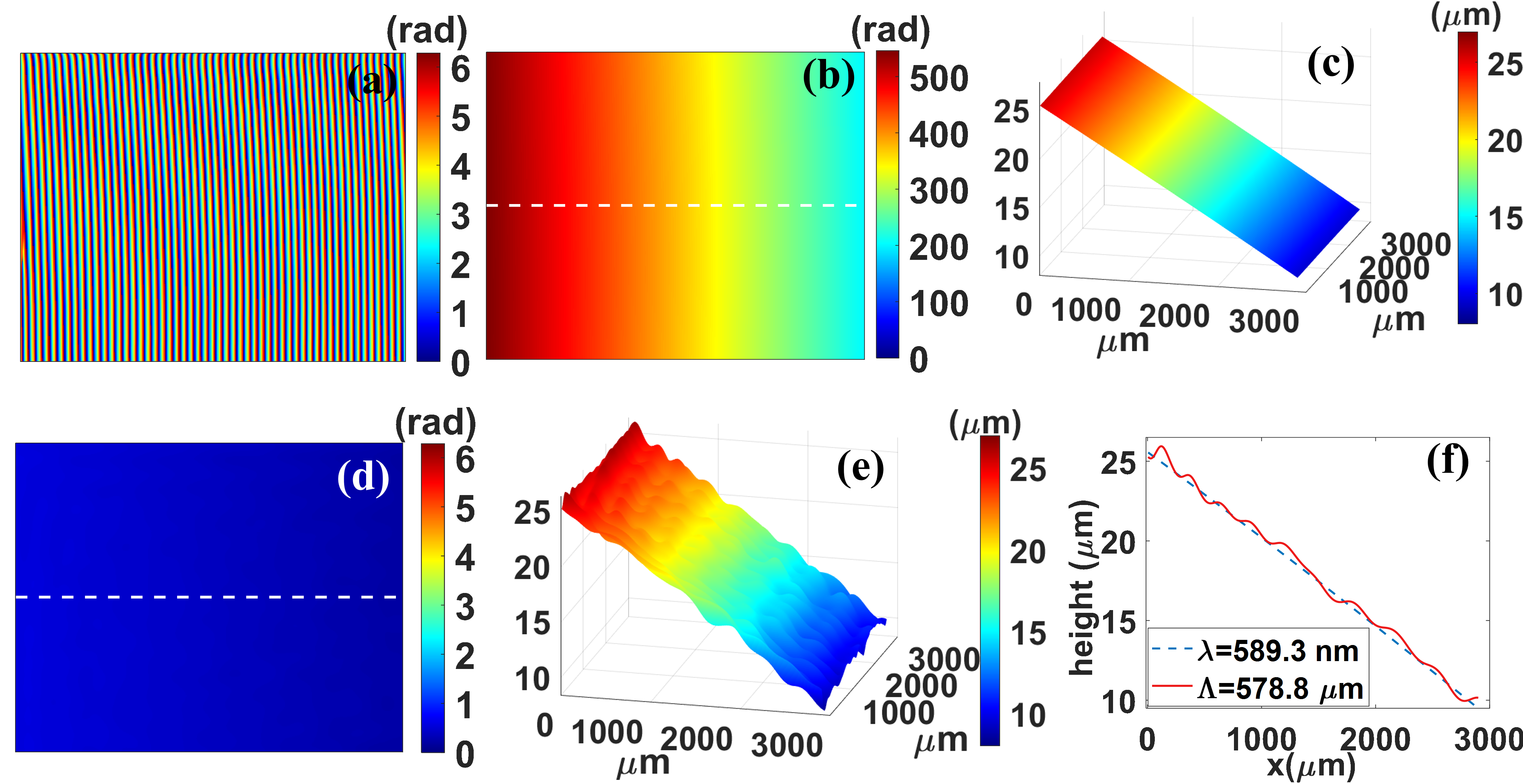}
	
	\caption{Measurement results. (a) Wrapped Phase, and (b) Unwrapped phase for $\lambda=589.3\ nm$. (d) Reconstructed phase for  $\Lambda=578.8\ \mu m$. (c) and (e) 3D representation of height. (f) Comparison of height profile along white lines in (b) and (d).}
	\label{Pic4}
\end{figure}

Figures \ref{Pic4}(a) and \ref{Pic4}(b) show two-dimensional (2D) wrapped and unwrapped phase maps for $\lambda=589.3\ nm$, respectively.
Figure \ref{Pic4}(d) presents the reconstructed 2D phase map using Eq. \ref{eq6} for $\Lambda=578.8\ \mu m$. It is clear in Figure \ref{Pic4}(d) that the phase range is less than $2\pi$
, and thus, the need for phase unwrapping algorithms is eliminated. While the maximum wrapped phase for the dual-wavelength method is 0.5 radians, the maximum unwrapped phase for the single-wavelength method is 546 radians, which is approximately 87 times \(2\pi\).
The optical path difference (OPD) can be calculated as $OPD=\Lambda\Delta\phi/2\pi$, and since the light travels twice in the Michelson interfrometer, the air wedge height distribution $h=\Lambda\Delta\phi/4\pi$. The 3D surface of the air wedge for $\lambda$ and $\Lambda$ is shown in Figs. \ref{Pic4}(c) and \ref{Pic4}(e), respectively.
To comparison one-dimensional (1D) profiles height of air wedge for $\lambda$ and $\Lambda$ along white dashed lines in Figs. \ref{Pic4}(b) and \ref{Pic4}(d), are shown in Fig. \ref{Pic4}(f).  The 1D profiles plotted for single-wavelength and dual-wavelength roughly match each other.

In another experiment, we shifted the movable mirror by the step of  $30\ \pm \ 4 \ \mu m$ to achieve different heights. 

\begin{figure}[h!]
	\centering
	\includegraphics[width=0.7\textwidth]{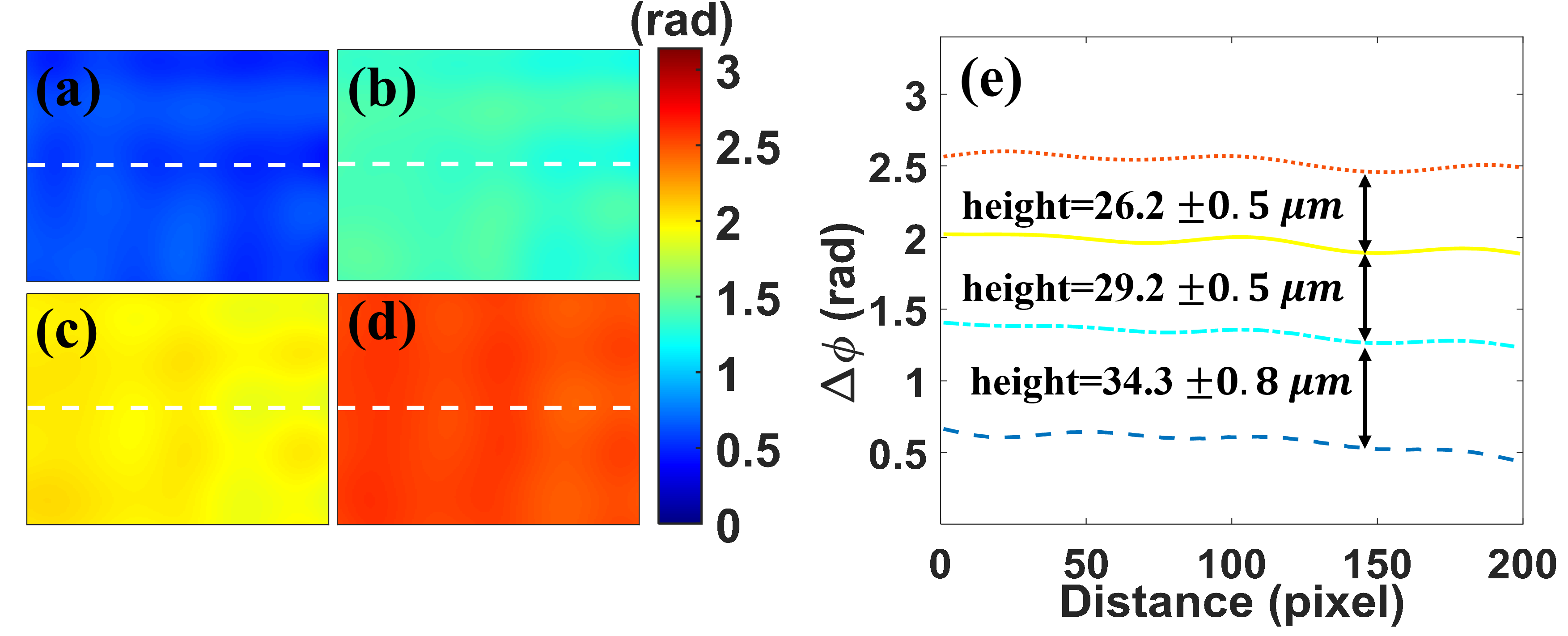}
	\vspace{-3pt}
	\caption{Measurement results. Reconstructed phase maps for heights difference (a) h=\ $30\ \mu m$, (b) h= $60\ \mu m$, (c) h= $90\ \mu m$, (d) h= $120\ \mu m$. (e) Phase difference profiles along the white dashed lines, correspond to samples with heights of 
		$30,\ 
		60, \
		90 \,$ and 
		$120 \ \mu m$, respectively, from bottom to top.}
	\label{Pic6}
\end{figure}

Figures \ref{Pic6}(a) - \ref{Pic6}(d) show the 2D phase maps extracted from the recorded holograms for heights difference of $30\ \mu m$, $60\ \mu m$, $90\ \mu m$, and $120\ \mu m$, respectively. Due to the large synthetic wavelength $\Lambda=578.8 \ \mu m$, the phase maps are obtained without ambiguity. Clearly, the phase difference increases uniformly with the increase of OPD. Figure \ref{Pic6}(e) shows 1D profiles of the phase difference along the white dashed lines in Figs. \ref{Pic6}(a) - \ref{Pic6}(d). By subtracting the heights obtained from the reconstructed phase differences of consecutive holograms, the step heights can be determined. The step heights between the curves are achieved $34.4\ \pm \  0.8 \ \mu m$ and $29.2\ \pm \  0.5 \ \mu m$, and $26.2\ \pm \  0.5 \ \mu m$, which closely matches the actual values.

\begin{figure}[h!]
	\centering
	\includegraphics[width=0.7\textwidth]{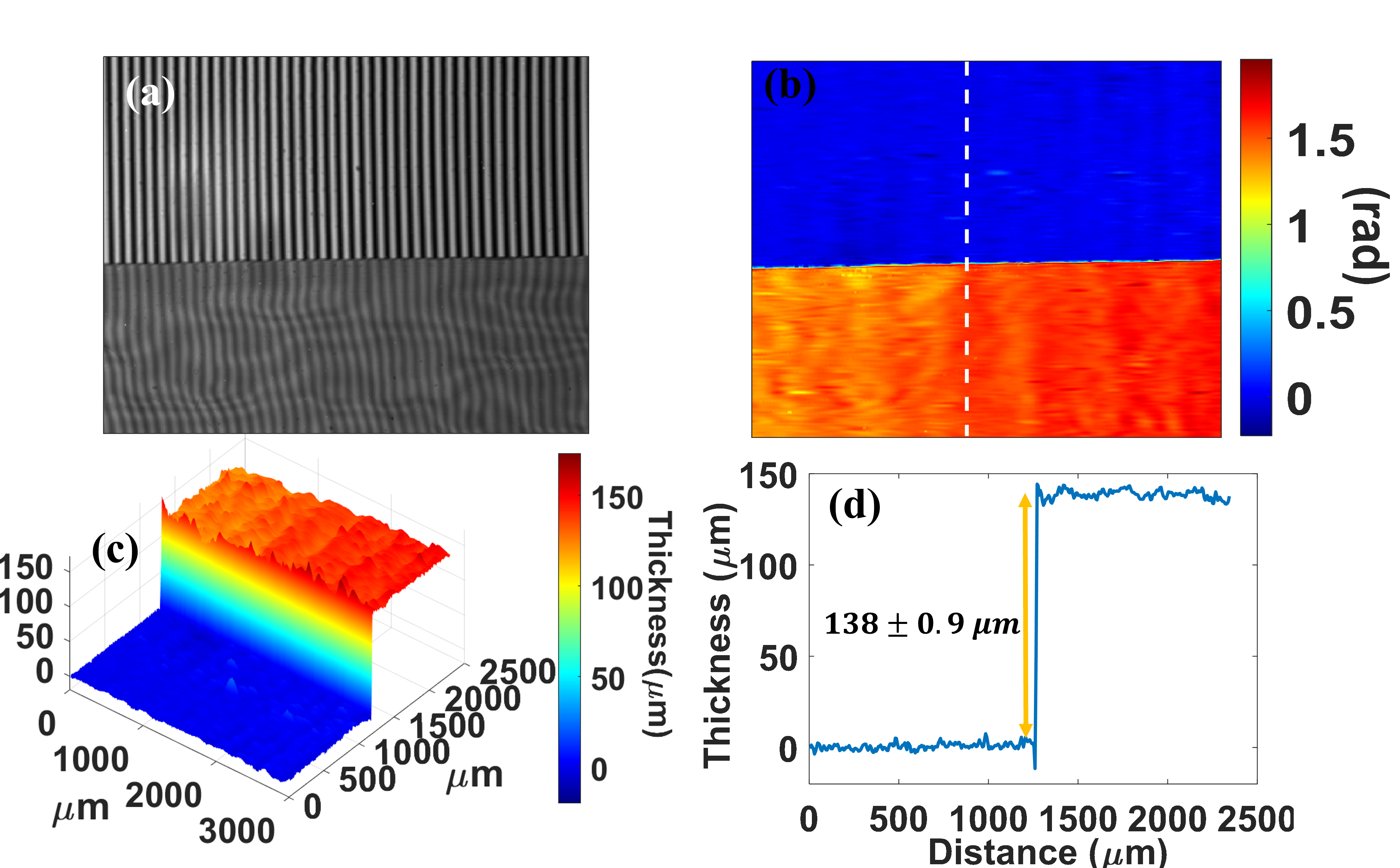}
	
	\caption{Measurement results. (a) Recorded hologram of the glass plate. (b) Reconstructed phase with proposed method. (c) 3D thickness of glass plate. (d) 1D thickness profile.}
	\label{Pic7}
\end{figure}

Also, based on our proposed method, we performed the measurements on a glass plate with a thickness of $140\ \pm 10 \mu m$ that is located in half of field of view in one arm of the interferometer as a step sample.
Figure \ref{Pic7}(a) shows the recorded hologram of the glass plate. Figure \ref{Pic7}(b) presents the 2D reconstructed phase map by our proposed dual-wavelength method. The thickness of the glass plate is obtained using $d=OPD/2\Delta n$, where $d$ is the thickness of the sample, $\Delta n$ denotes the refractive index difference between the glass plate ($n_g=1.52$) and air ($n_A=1$). Figure \ref{Pic7}(c) shows the 3D thickness of the glass plate. Figure \ref{Pic7}(d) presents the 1D thickness profile of the sample along the white dashed line in Fig. \ref{Pic7}(b). The step height of about $138\pm 0.9  \ \mu m$ is obtained, which is consistent with the thickness of the plate.

\section{Conclusion}
In summary, We have presented a novel direct method for single-shot phase difference retrieval based on dual-wavelength off-axis digital holography. Phase reconstruction without calculating the phases of each wavelength separately enables the use of a sodium vapor lamp with two wavelengths to achieve the synthetic wavelength of about $0.6 \ mm$ , which is not feasible with traditional dual-wavelength methods. This capability allows the two wavelengths in the object and reference arms to pass a common path, making the setup compatible with a conventional and common-path single-wavelength configurations. The feasibility of the proposed method is confirmed by experimental results on an air wedge,  $30 \ \mu m$ step height differences, and a glass plate.

 \subsection*{Acknowledgements}
This work is based upon research funded by Iran National Science Foundation (INSF) under project No. 4027765.

\appendix


\end{document}